\documentclass[sigconf]{acmart}

\AtBeginDocument{%
  \providecommand\BibTeX{{%
    \normalfont B\kern-0.5em{\scshape i\kern-0.25em b}\kern-0.8em\TeX}}}

\copyrightyear{2021}
\acmYear{2021}
\setcopyright{rightsretained}
\acmConference[K-CAP '21]{Proceedings of the 11th Knowledge Capture Conference}{December 2--3, 2021}{Virtual Event, USA}
\acmBooktitle{Proceedings of the 11th Knowledge Capture Conference (K-CAP '21), December 2--3, 2021, Virtual Event, USA}

\acmDOI{10.1145/3460210.3493548}
\acmISBN{978-1-4503-8457-5/21/12}

\usepackage{balance}
\usepackage{graphicx}
\usepackage{longtable}
\usepackage{appendix}
\usepackage{array}
\usepackage{xcolor}
\usepackage{booktabs}
\usepackage{listings}
\usepackage{adjustbox}
\usepackage{dirtytalk}
\usepackage{float}
\usepackage{pdfpages}
\usepackage{multirow}
\usepackage[normalem]{ulem}
\usepackage{ragged2e}
\usepackage{todonotes}

\begin{document}
\fancyhead{}
\title{The Wind in Our Sails: Developing a Reusable and Maintainable Dutch Maritime History Knowledge Graph}
\author{Stijn Schouten}
\author{Victor de Boer}
\affiliation{%
  \institution{Vrije Universiteit Amsterdam\\Department of Computer Science}
  \streetaddress{De Boelelaan 1111}
  \city{Amsterdam}
  \country{The Netherlands}
}\email{{stijnscho@gmail.com, v.de.boer@vu.nl}}

\author{Lodewijk Petram}
\affiliation{%
  \institution{Huygens Instituut}
  \streetaddress{Oudezijds Achterburgwal 185}
  \city{Amsterdam}
  \country{the Netherlands}}
\email{lodewijk.petram@huygens.knaw.nl}

\author{Marieke van Erp}
\affiliation{%
  \institution{KNAW Humanities Cluster\\DHLab}
  \streetaddress{Oudezijds Achterburgwal 185}
  \city{Amsterdam}
  \country{the Netherlands}}
\email{marieke.van.erp@dh.huc.knaw.nl}

\renewcommand{\shortauthors}{Schouten et al.}

\begin{abstract}
Digital sources are more prevalent than ever but effectively using them can be challenging. One core challenge is that digitized sources are often distributed, thus forcing researchers to spend time collecting, interpreting, and aligning different sources. A knowledge graph can accelerate research by providing a single connected source of truth that humans and machines can query. During two design-test cycles, we convert four data sets from the historical maritime domain into a knowledge graph. The focus during these cycles is on creating a sustainable and usable approach that can be adopted in other linked data conversion efforts. Furthermore, our knowledge graph is available for maritime historians and other interested users to investigate the daily business of the Dutch East India Company through a unified portal.


\end{abstract}

\ccsdesc[500]{Applied computing~Arts and humanities}
\ccsdesc[500]{Information systems~Graph-based database models}

\keywords{Knowledge Graph, Digital Humanities, Maritime history}


\maketitle

\section{Introduction}
Recent large-scale digitization efforts have increased the volume of digitally available sources enormously~\cite{claire_warwick_digital_2012}. This is a huge boon to digital humanities research, a multidisciplinary field, where researchers utilize computational processing power to develop and present insights that were not possible before in traditional humanities research~\cite{berry_understanding_2012}.  However, using these digitized sources in large-scale research is not always straightforward. When data sets are available as standalone files, researchers have to go through the laborious process of collecting the different files, interpreting and refining the data, defining relations between different entities, and finally creating the queries and filtering operations needed to answer their research questions. This can be a complex process; for one thing, interpretation of variables may differ between researchers~\cite{haslhofer_knowledge_2018} and the process is often repeated by every new researcher who wants to work with those data sets. 

In this project, we create a knowledge graph of data related to the Dutch East India Company (Vereenigde Oostindische Compagnie, abbreviated as \textit{VOC}). The VOC is often considered the first multinational company. It was a large Dutch trading company that was founded in 1602 and remained in business until the end of the eighteenth century~\cite{gelderblom_formative_2013}. The VOC kept records of its ships, their cargo, and the sailors and soldiers who ventured to Asia. This information is contained in four large data sets, three of which are available online from the websites of the Dutch National Archive (VOC Opvarenden, or VOC Crew members) and the Huygens Institute (Dutch Asiatic Shipping, Bookkeeper General Batavia and Places). The last data set was provided to us by the researchers from the Huygens Institute. The datasets are described in more detail in Section~\ref{sec:firstiteration}. 
Making these data sets available in a well-structured knowledge graph allows digital humanities scholars to accelerate research with these rich and unique data~\cite{hughes_evaluating_2012}. While many humanities researchers work with structured data such as databases and spreadsheets, linked data is still uncharted territory for many. This is partly due to unfamiliarity with the technology, and its at times steep learning curve to start using it~\cite{blanke2012linked}. 

The goal of this project is to design and develop a usable and sustainable knowledge graph, that will be hosted by the Huygens Institute, the research institute that manages three of the four data sets used in this project, and that aims to drive data-driven research. We adopt an iterative design process; in each iteration we capture additional knowledge from domain experts and the data sets. We are not the first to create linked data for historical resources (cf.~\cite{blanke2012linked,hyvonen2020linked}) or even the first to do this for the historical maritime domain (cf.~\cite{de_boer_dutch_2015}) However, our work is the first to focus on sustainability of the knowledge graph, as well as in bringing together four Dutch historical maritime data sets.  

The challenges in our design process are that the approach should be sustainable and that the designed artifact should be useful for researchers and domain experts. Measuring and determining the most sustainable approach is not straightforward and can be subjective. We therefore define four quality attributes that can be used to measure, steer, and lead the design to a sustainable solution. These design requirements are set in cooperation with the project stakeholders and are as follows:

\begin{itemize}
	\item \emph{Transparency:} Clear and well-described design decisions and rules.
	\item \emph{Accessability:} Results and intermediate results should be accessible.
	\item \emph{Re-usability:} Re-usable methods and semantic models.
	\item \emph{Maintability:} Future-proof knowledge graph and infrastructure.
\end{itemize}

\par \noindent We evaluate our knowledge graph using competency questions following~\cite{azzaoui_scientific_2013}.


\section{Related work}

This project is on the intersection of computer science and humanities, therefore we discuss relevant related work from both disciplines, starting with best practices in knowledge modelling from the computer science/semantic web domain. 

The design guidelines of developing an accessible, transparent, reusable, and maintainable knowledge graph are closely related to the FAIR~\cite{wilkinson_fair_2016} principles. Waagmeester et al.~\cite{waagmeester_wikidata_2020} provide us with tips and pointers on how to develop a knowledge graph according to the FAIR principles. Dimou~\cite{janev_knowledge_2020} dives more into the technical side of knowledge graph engineering and inspires further research and comparison of these approaches. Finally, Jovanovik~\cite{Jovanovik} brings these together in his linked data application development methodology, which focuses both on FAIR principles and the technical implementation. 

Our work is not the first to design a knowledge graph for the historical maritime domain. 
De Boer et al.~\cite{de_boer_dutch_2015} started in 2014 with Dutch Ships and Sailors (DSS). The researchers involved in this project connected four different data sets in a first effort to unify the scattered Dutch historical maritime data. Additional work in this domain was carried out by Entjes~\cite{entjes_linking_2015} in 2015.
Both projects score high in terms of \emph{transparency} and \emph{accessibility}, with design and conversion decisions being documented and reusable mapping scripts made available. 
In terms of re-usability, the verdict is less clear. DSS uses elements from multiple general-purpose ontologies and data models, and combines these with concepts and classes from a specific DSS model to describe the maritime activities. This design decision has definite upsides as it is generally recommended to apply widely used models and ontologies to enable interoperability with other graphs on the web.\footnote{\url{https://www.w3.org/standards/semanticweb/ontology}} The re-use potential within the (maritime) history domain is low, however: it is unlikely that the \textit{dss:Ship} concept is used between projects outside of the DSS sphere.

As a use case for the OntoME platform Beretta et al.~\cite{beretta1} developed another ontology for Dutch maritime data on ships and sailors in 2019.\footnote{\url{http://ontome.net}} OntoME is aimed at maintaining and distributing ontologies for digital heritage projects, and the approach of Beretta et al. therefore scores high in terms of maintainability. Furthermore, their ontology is based on the CIDOC CRM~\cite{cidoc2003cidoc}, a widely accepted and highly detailed model for digital cultural heritage. Concepts in this model, such as voyages, have a better chance of being re-used in other semantic web projects, thus offering great alignment possibilities. It is accessible and available for re-use on the OntoME platform.\footnote{\url{http://ontome.net/project/28}} The transparency and accessibility design principles of Beretta et al. and, more generally, the OntoME platform, leave room for improvement. Design decisions are not openly published, making it difficult for outsiders to understand and follow the thought process during the design.

Each of the discussed projects has its strengths and weaknesses. We adapt some of their strong aspects in our project, such as the transparent methodology of DSS and the model maintainability of OntoME. 

\section{Research method and design}

\subsection{Design decisions}
In this project, the CIDOC CRM is selected as the ontology of choice. In contrast to the aforementioned DSS project, this will allow for further interoperability with other Digital Humanities Linked Data solutions. CIDOC CRM is a de facto domain standard and a safe choice; as it is for example also already used in the OntoME project. 

We use GraphDB for data conversion.\footnote{\url{https://graphdb.ontotext.com/documentation/standard/release-notes.html}} GraphDB is transparent due to its interface; users can directly intuitively influence the mapping. It is reusable due to the ability to import and export mapping tables between projects. Maintainability also scores high due to the active development of the application. Other conversion tools were tested and examined as well. These include but are not limited to the tools of the Datalagend ecosphere\footnote{\url{https://iisg.amsterdam/en/clariah}} which we found to be well capable of quickly producing high volumes of RDF data with minimal user input, but too rigid for modelling and linking entities. The RML-based editors, we investigated such as RMLEditor\footnote{\url{https://rml.io/tools/rmleditor/}} places a limit on file size and requests for an extension were not answered. Other RML editors (such as CARML,\footnote{\url{https://github.com/carml/carml}}, and Yarrrml (matey)\footnote{\url{https://rml.io/yarrrml/matey/}})  lack a workable visual or GUI representation. This downside makes the recommendation to use such an approach difficult for users outside computer and information sciences.  Therefore, we deemed GraphDB the most suitable for our research goals.

\subsection{Iterative design}
We select \textit{design science} as our central design methodology, as the primary goal of our project is the development of an artifact: the knowledge graph. In Design Science research, several research methods and frameworks exist (cf. \cite{wieringa2014design,hevner_design_2004}). The main takeaway of these theories is the design-test cycle. In this iterative cycle, the researcher uses the designed artifact to solve the given problem while grounding the artifact in a socio-technical foundation. The artifact is evaluated and the evaluation results are used in the second iteration of the design-test cycle. The cycle continues until a certain predefined threshold is met. 

We complete two design cycles. This allows for a review and evaluation period while still being time-efficient. In the first design-test cycle, we implement the approach of Jovanovik~\cite{Jovanovik} with the goal of re-using some of the steps during the second design-test cycle. During the first design-test cycle we also conduct an open and unstructured interview with a domain expert from the Huygens Institute. The domain expert is responsible for some and involved in most of the curation efforts of data sets used during in this project. An interview does not only help with understanding the data sets but also creates an understanding of the use case, how domain experts plan to use the artifact, and what they expect from it. Not all questions can be asked and answered during a single interview; some questions may arise during the cycle. Due to the pandemic, in person collaborations at the research institute's office were not possible; we therefore keep close contact via email. Finally, the documentation provided by the authors of the previous historical maritime conversions (\cite{de_boer_dutch_2015,entjes_linking_2015,beretta1}) is examined during desk research.

The first review period has two goals. First, errors, shortcomings, and other issues are investigated. Second, the review period enables the start and direction of the second design iteration. An unstructured discussion session with digital humanities experts and the domain expert completes the review period.

Finally, in the second review period, we test the design again and conduct a workshop with several stakeholders from the Huygens Institute and the University of Amsterdam. In this workshop, the developed artifact can be user-tested, thus validating the research results. Aside from the workshop, a short data story is written to illustrate the usefulness of the created knowledge graph (presented in Section~\ref{see:evaluation}).

We evaluate the \emph{sustainability} aspect of our project through the design guidelines. \emph{Usefulness} can be measured with the completion of the competency questions~\cite{azzaoui_scientific_2013}. Our competency questions come from two sources: the domain expert of the Huygens Institute and a related project\cite{entjes_linking_2015}. In the list of competency questions below, (D) indicates that the domain expert submitted the question; questions derived from the related project are denoted with (R):

\begin{itemize}
    \item (D) Which VOC Chamber was accountable for the largest number of slaves transported?
    \item (D) How were the shipping routes in Asia divided between the various VOC chambers (for example, did ships from a specific chamber predominantly sail on certain routes)?
    \item (D) What was the average value of cargo on VOC return voyages per crew member/ship's ton, and how did this evolve over time?
    \item (D) To what extent was the value per ships' ton on return voyages correlated with the skipper's track record (how much experience, i.e., how many previous voyages / how quickly did a ship get to Asia on an outbound voyage)?
    \item (R) How did the price of goods change over time?
    \item (R) Perform a search for specific information in large data sets, such as names;
    \item (R) Can external factors, such as war, be linked to the intensity of shipping?
\end{itemize}

\section{First iteration}\label{sec:firstiteration}
As discussed in the methodology section, we derive the structure of the iterations from the work of Jovanonik~\cite{Jovanovik}. 

\begin{figure*}[ht!]
\includegraphics[width = .9\textwidth]{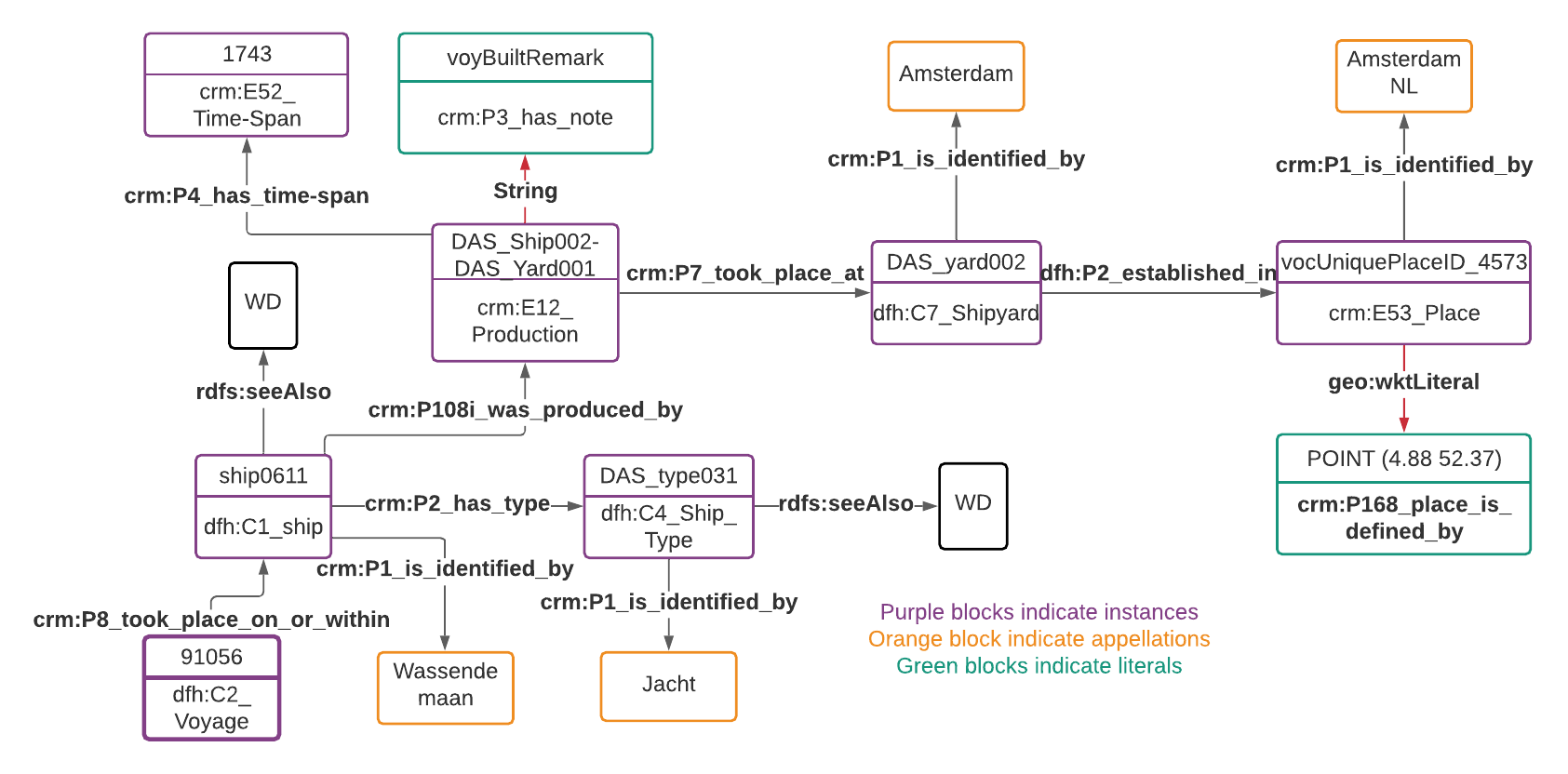}
\caption{Example of mapping branch}
\label{Fig7}
\end{figure*}

\subsection{Data sets}
We include three VOC related data sets in the first design iteration:

\begin{itemize}
    \item \emph{Dutch-Asiatic Shipping (DAS)} This data set contains information about all voyages from the Netherlands to Asia and back made by the VOC and its predecessors. Some core details are available for every voyage, such as the VOC Chamber that outfitted and contracted the ship, and the places and dates of departure and arrival.\footnote{\url{http://resources.huygens.knaw.nl/das} Retrieved on: 13-05-2021} During these voyages, events occurred, such as a storm or a capture of an enemy vessel. These events are stored in the voyParticulars column. Voyages had a master and were made by ships that have a name, type and tonnage, and were built at shipyards. Ships could make multiple journeys and sometimes have their name changed. DAS contains the information of 4,700 voyages from the Netherlands to Asia and of 3,400 return voyages. 
    \item \emph{Bookkeeper-General Batavia (BGB)} In Batavia (now Jakarta in Indonesia), the bookkeeper and his clerks kept track of all imports and exports over sea. For a sizable portion of the eighteenth century, these records have survived. They were entered into a database in the past decade and published online.\footnote{\url{https://bgb.huygens.knaw.nl} Retrieved on: 13-05-2021} Similarly to the voyages in DAS, voyages in BGB leave and arrive on a specific date to and from a specific port. There are also overlapping voyages between the different data sets. These overlapping voyages were identified by previous research. The voyages transported several goods with a specific value, unit of measurement, and specification. Voyages partook on ships that have different names over time. Note that in comparison to DAS, less information about ships is captured in this data set. BGB consists of 55 volumes of the 18th century containing 18,722 records. 
    \item \emph{Places} The Places data set is a single table containing information about geographical places related to DAS and BGB. It contains toponym attestations from DAS and BGB that are linked to a standardized place name (toponym followed by a space and the ISO 3166-1 alpha 2 country code). GeoNames and Wikidata URIs, as well as geographical coordinates are available.\footnote{\url{https://www.geonames.org/}; \url{https://www.wikidata.org/}} The places dataset consists 4,252 instances. 
\end{itemize}

\subsection{Data and ontology modeling}
In the data and ontology modeling phase, we aim to map data concepts to our base ontology (CIDOC CRM) and add new attributes to that model if one of these concepts does not fit the model.

We initially identify four main components: voyages, ships, persons, and companies. Their relations include, but are not limited to, a voyage taking place on a ship, a sailor participating on a voyage, and the VOC chamber outfitting the ship. With these high-level concepts defined, we can specify the lower-level concepts.

Voyages form the starting point of the lower-level, detailed concepts. A voyage is the journey that a ship undertakes, moving from one place to another. In this journey, sailors, soldiers, and other personnel are employed to ensure that the ship's cargo safely reaches its destination. 

We aim to re-use CIDOC CRM classes and properties where possible. If no good fit is available, we re-use classes and properties of Beretta et al.~\cite{beretta1}. Finally, we create new classes and properties accommodate the envisioned data model. These newly created properties and classes are integrated into the CIDOC CRM model as a subclass of existing classes, thus making them interoperable with other CIDOC CRM extensions and models.

For reasons of readability, we do not display the entire ontology model, but instead show a branch starting from a voyage in Figure \ref{Fig7}. The complete mapping figures can be found on the GitHub.\footnote{\url{https://github.com/stijnschout3n/thewindinoursails}}

We initially defined the \textbf{voyage} instance as a subclass of \textit{CRM:E7 Activity}. We later changed this definition to a subclass of \textit{CRM:E9 Move} to accommodate the linking of goods (physical objects) to the voyage via the property \textit{CRM:P25 moved}. We define \textbf{ships} as a subclass of \textit{CRM:E22 man-made object}. The property \textit{CRM:P8 took place on or within} is used as property between the \textbf{voyages} and the \textbf{ships}. (Incidentally, the property description of \textit{CRM:P8 took place on or within}, even uses this relation as an example: `It describes a period that can be located with respect to the space defined by an \textit{CRM: E19 Physical Object} such as a ship or a building.') We define the \textbf{name of the ship} through the property \textit{CRM:P1 is identified by}, which links the ship instance to a \textit{CRM:E41 Appellation} instance. Other definitions and classifications, such as ship types and productions, can be found in Figure \ref{Fig7}.

\subsection{Linked Data Transformation}
In the transformation phase, we transform the acquired knowledge into RDF. In short, a table structure is mapped to the different positions within a RDF triple. We summarize our most important design decisions:

\begin{description}
    \item [Time] Time plays a central role in the maritime model. It relates to many concepts. Such as the arrival or departure of ships. In the initial phases of the conversion, the dates were sanitized and formatted according to the ISO 8601-1:2019 standard, which does not allow for uncertainties. The reasoning behind this decision is as follows: more complex date formats are difficult to calculate with and can therefore lead to lower usability of the artifact. In consultation with the domain expert, we decided to add the original dates (which sometimes state uncertainties) as an extra attribute for researchers to consult while retaining data usability. Dates were formatted, according to CRM standards, as URIs and not as XSD:Date literals. This decision helps link events together that happened during the same time-span. 
    \item [URIs] We create URIs for unique (and unambiguous) entities. Identifying entities across different data sets is rather tricky\cite{christen_data_2012} and it is not captured in this research project. Fortunately, other researchers have performed data resolution and linking on these data sets \cite{hendriks_recognising_2020,petram_data_nodate}. For example, the BGB data set is linked at some entries to DAS voyages. In these cases, the DAS voyage entity is reused, effectively linking the two data sets. For goods shipped by the VOC, we create unique URIs for every good that was carried on a specific voyage. This is to allow for linking voyage-specific information (such as quantity or price) to the good.
\end{description}

\subsection{Data interaction}\label{sec:datainteraction}
After converting and connecting the different data sets, we make our knowledge graph accessible. We do this such that researchers are able to query the created knowledge graph and wrangle the retrieved data to achieve their research goal. Our knowledge graph can be accessed via \url{https://dutchshipsandsailors.nl/}.

During the project, we also experimented with the Metaphacts platform\footnote{\url{https://metaphacts.com/product}} which is well integrated with GraphDB and comes in a pre-configured Docker container. Second, the Metaphacts platform hosts very usable graph exploration tools. These tools can help in answering the defined competency questions. Due to licensing issue, we cannot expose our data through Metaphacts, but invite readers to try out this platform, as well as exploring the data through the triplestore of their choice. 




\begin{figure*}[ht]
\includegraphics[width=\textwidth]{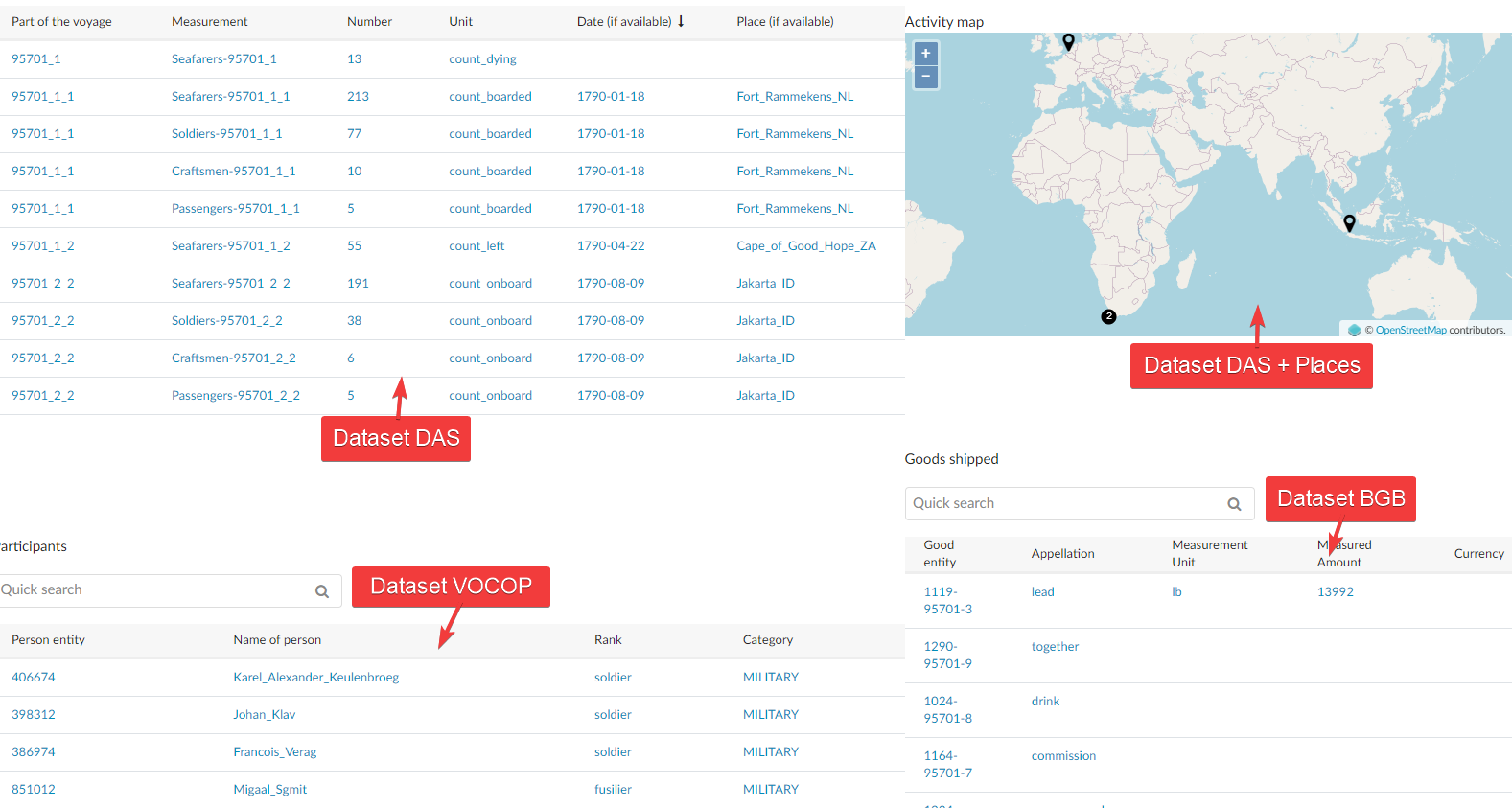}
\caption{Visual integration of the four data sets}
\label{fig:visualintegration}
\end{figure*}

\subsection{Review of the initial design}
Modeling data sets into an existing ontology is a delicate process, even with high-level ontologies, such as CIDOC CRM. There is a continuous push and pull between modeling from a data perspective and modeling from an ontology perspective. In a perfect world, the ontology and the data are a match. In this case, there exists a class and property for every envisioned concept. Nevertheless, this is not the case when using a general ontology, as it defeats its purpose. Therefore, the envisioned concepts should be fitted within the scope of the ontology. 

Differences in semantics between classes is not unique to this project. The domain experts foresaw this and started developing an extension for the CRM ontology. This extension, focusing on social, economic, and legal life, allows for modeling concepts outside of the museum artifact realm.\footnote{\url{http://ontome.net/project/64} Last visited: 16-09-2021} 

We now take a step back to a more meta-level by checking the research questions' progress. Any forthcoming discrepancies can be fixed in the second design iteration. 

\textbf{Usability:} The solution is approaching usable compared to the research goals. In Subsection~\ref{sec:datainteraction}, some of the competency questions could be answered with the interface. However, not all competency questions can be answered at this moment. The first design iteration did not include the very extensive VOC Opvarenden (VOCOP) data set. This data set contains rich records of the sailors and soldiers who crewed the VOC ships. These records will be included in the second design iteration, thus fulfilling the \textit{usability} requirement. 

\textbf{Sustainability:} This is not as strictly defined as the usability requirement. We can only review the approach and the decisions made during the first design iteration and determine where the sustainability can be improved. The approach, thus far, is \emph{transparent} due to the extensive documentation. The triple store allows for an \emph{accessible} environment where the knowledge graph can be accessed interactively. By reviewing different approaches and methods, the current approach seems reasonably \emph{reusable} as well. As we use a CRM extension in the second iteration, we are actively promoting reusability.

\section{Second iteration}
In the second design iteration, we process the results of the first review period, and try to resolve any shortcomings regarding the research questions. This iteration will be noticeably compacter due to our improved understanding of the source data, making mere improvements rather than additions, and having fewer data sets to convert.

The design thus far consists of voyages (DAS), cargo moved on voyages (BGB), and the places where the voyages called (Places). However, several competency questions are related to the people involved in these activities. In the second design iteration, we add persons to the design to solve these competency questions.

The VOC kept extensive records of its contracted crew members (`Opvarenden' in Dutch). For the late seventeenth and eighteenth centuries, many pay ledgers have been preserved. Pay ledger entries include personal data, ranks, wages, and references to voyages. The voyages overlap with the voyages of the DAS data set. The information of 774,200 crew members was preserved in the VOC Opvarenden data set.  

The core of the model does not change in the second iteration of the design. The changes in the second iteration are more apparent in the details. One of the more critical changes is related to the structure of the voyages. In summary of the previous iteration: a \textbf{voyage} is linked to a starting point via a departing activity and ends with the arrival of the voyage at its destination. Persons and goods are directly linked to the voyage; we assume that they were involved during the whole duration of the voyage. However, in the VOCOP data set, this is not the case. Persons could change ships during a stopover, effectively participating in two voyages. The first iteration could not adequately convey this concept.

To correctly convey this concept, we introduce a new class. The \textbf{leg} class is, similar to the voyage, a \textbf{phase}. It indicates a continuous part of the voyage from and to a port of call. A voyage can consist of multiple legs. The time between the end of the first leg and the beginning of the second leg is the stopover. Finally, the sailors are linked to the departing and arriving properties to indicate (dis)embarkment activities. 

We choose the CIDOC CRM extension for economic, social, and legal life (SDHSS prefix) for mapping the VOCOP data set. To relate persons to a voyage, the \emph{participation} concept is used. Other mappings include ranks and their descriptions.

The changes made during the second design iteration aim to improve the \emph{reusability} of the artifact by utilizing an extended ontology. Creating mapping figures and tables improves \emph{transparency} concerns. Converting the VOCOP data set ensures that the \emph{usability} aspect as defined by the competency questions is met. To improve the \emph{maintainability} aspect of the artifact, the OntoME platform is proposed. The power of OntoME is in the reuse of ontologies. The application has similarities with software development platforms such as Github in that it allows users to collaborate and discuss aspects concerning the creation and utilization of classes and properties. Furthermore, other researchers can easily reuse the captured knowledge within the ontology for their own projects. 

Changes in the modeling have a cascading effect on the front-end part of the artifact. In the second design iteration, these issues are examined and fixed. Furthermore, the VOCOP data set was also integrated in the knowledge graph. An example of how the different sources can be explored in a single screen through the Metaphacts platform is shown in Figure~\ref{fig:visualintegration}.

Community interaction is not necessarily a research goal. However, it can have clear benefits. As \cite{waagmeester_wikidata_2020} points out, sustainable growth of the knowledge graph can be realized with community involvement. A proposal that captures this issue is the utilization of the Geovistory platform.\footnote{\url{https://docs.geovistory.com/}} This platform can interact with a knowledge graph and hosts a text editor that can recognize entities and propose linkage to existing entities within the knowledge graph. Academics and citizen-scientists could for example, with approval, enrich the knowledge graph by browsing through the available data and adding connections within the graph.

\section{Evaluation}\label{see:evaluation}
With the completion of the second design iteration, most of the issues discovered during the review period are resolved. These include, but are not limited to, better utilization of the CIDOC CRM model and its extensions, ontology management, community involvement, and the conversion of data on crew member (VOCOP). In the second review, we explore these improvements and related them to the primary research goal. The review consists of a workshop with five domain experts from the Huygens Institute and the University of Amsterdam. The expertise of the domain experts covered maritime history, digital data management and data science, all with an affinity for (digital) humanities research.

In the workshop, multiple discussions where brought up to challenge design decisions made during the research project. These topics include, but are not limited to, the utilization of CIDOC CRM, the conceptual mapping of a voyage, and the proposed infrastructure. The feedback was overall positive  but also provided constructive criticism and included recommendations to put more focus on data provenance in future iterations. 


Initially, one of the goals of this project was to create a usable knowledge graph in the field of Dutch historical maritime data. Discussion on what makes a knowledge graph useful stems mainly from the perspective of maritime historians, as they are the primary stakeholders. They provided the competency questions which are used to test and improve the artifact. The answer to the solvability of these questions can be found in Table~\ref{tab:competencyquestions}. 

However, the developed artifact is not only intended to be used by maritime historians. Other interested parties can be drawn in by their research of an event or a long-lost forefather who had a remarkable VOC career. Determining if the knowledge graph is useful for these users is challenging as there is no clear picture of who these people are and what their research questions entail. Nevertheless, we found one subject and their research question is explored in a short data story.\footnote{\url{https://github.com/stijnschout3n/thewindinoursails/blob/main/Datastory.pdf}} In short, the data story briefly dives into the life of a sailor, Hendrik Prins. Quantitative and qualitative data analytics are performed in order to portray the personal life and career of Hendrik Prins. The developed knowledge graph makes the process of combining data sources straightforward and quick.

\begin{table*}[ht]
\centering
\begin{tabular}{ll} 
\toprule
Question                                                         & Answer                                                                                                                                                                                 \\ 
\hline
Which VOC Chamber was accountable                                & \multirow{2}{*}{\begin{tabular}[c]{@{}l@{}}Can be answered by querying voyages \\that transported slaves\end{tabular}}                                                                 \\
for the highest number of slaves transported?                    &                                                                                                                                                                                        \\ 
\hline
How were the shipping routes in Asia
  divided between~          & \multirow{3}{*}{\begin{tabular}[c]{@{}l@{}}Can be answered. Already a preconfigured \\component in Metaphacts\end{tabular}}                                                            \\
the various VOC chambers (for example, did ships                 &                                                                                                                                                                                        \\
from a specific chamber only sailed on certain routes)?          &                                                                                                                                                                                        \\ 
\hline
What was the average value of cargo on                           & \multirow{3}{*}{\begin{tabular}[c]{@{}l@{}}Can be answered. Goods, \\participants and prices are available\end{tabular}}                                                               \\
VOC return voyages per crew member/ship's ton,                   &                                                                                                                                                                                        \\
and how did this evolve over time?                               &                                                                                                                                                                                        \\ 
\hline
To what extent was the value per ships'
  ton on return~         & \multirow{4}{*}{\begin{tabular}[c]{@{}l@{}}Can be answered with data research,\\the components are available in the~\\knowledge graph. Data paper provides\\an example.\end{tabular}}  \\
voyages correlated with the skipper's track record~ ~~           &                                                                                                                                                                                        \\
(how much experience, i.e., how many previous voyages~           &                                                                                                                                                                                        \\
/ how quickly did a ship get to Asia on an outbound voyage)?     &                                                                                                                                                                                        \\ 
\hline
\multirow{2}{*}{How did the price of goods changed over
  time?} & Can be answered, goods are linked                                                                                                                                                      \\
                                                                 & to~voyages which have dates                                                                                                                                                            \\ 
\hline
Search for specific information in~                              & \multirow{2}{*}{\begin{tabular}[c]{@{}l@{}}Can be answered with search\\~components of Metaphacts\end{tabular}}                                                                        \\
large data sets, such as names.                                   &                                                                                                                                                                                        \\ 
\hline
Can external factors, such as war,~                              & \multirow{2}{*}{\begin{tabular}[c]{@{}l@{}}If the external factors are captured in the \\same ontology, then yes.\end{tabular}}                                                        \\
be linked to the intensity of shipping?                          &                                                                                                                                                                                        \\
\bottomrule
\end{tabular}
\caption{Competency questions including statement regarding solvability}
\label{tab:competencyquestions}
\end{table*}

\section{Discussion and Conclusion}
In this paper, we presented a reusable and maintainable knowledge graph for Dutch maritime history. The knowledge graph was created using an iterative design and development process that involved the conversion and linking of four different data sets concerning the Dutch East India Company's business dealings. The resulting knowledge graph consists of more than 21 Million RDF triples. The dataset is available through an online triple store\footnote{\url{https://semanticweb.cs.vu.nl/eots/}} and can be queried through a SPARQL endpoint\footnote{\url{https://semanticweb.cs.vu.nl/eots/sparql}}. The data itself is made available under open licences\footnote{The Knowledge Graph and scripts to produce it is published as four separate datasets through the following DOIs: Places dataset: \url{https://doi.org/10.5281/zenodo.5509814} ; VOCOPV dataset: \url{https://doi.org/10.5281/zenodo.5509820} DAS dataset: \url{https://doi.org/10.5281/zenodo.5507139} ; BGB data \url{https://doi.org/10.5281/zenodo.5507126}}. The data model and knowledge graph were evaluated using competency questions provided specifically for this project by a domain expert as well as competency questions formulated in earlier work for this domain. Furthermore, the model and knowledge graph were reviewed by five different experts with complementary knowledge regarding the domain and computational analysis in the context of digital humanities research. 

The path for future work, appending more data sets and improving the model, is paved by the used methodology. By following the framework of Jovanovik~\cite{Jovanovik}, which focused on reusable components, other researchers can complete further iterations. Our  documentation is made available to aid researchers in understanding the decisions made during the design. Other data sets could include the Northern Musterrols\footnote{\url{https://www.noordelijkscheepvaartmuseum.nl/nl/collectie-en-onderzoek/onderzoek/monsterrollen}} or the Archangel and Elbing voyages\footnote{\url{http://resources.huygens.knaw.nl/archangel/app/inleiding}} in the Baltic sea and similar data sets from other countries. 

It should also be noted that even with many changes in the second design iteration, the processing time of the second design iteration was much shorter than the first iteration. Whereas the first iteration took around six weeks to complete, we finished the second design iteration in two weeks. There is a steep learning curve in understanding the concepts, tooling, and methods used. Hopefully, other researchers will initiate new iterations and utilize the results of this research project to face a less steep learning curve. 

There are limitations to our project. In the second design iteration, the proposed infrastructure is expanded with the use of OntoME and Geovistory. However, the two extensions are not integrated entirely as they are web-based and, therefore, difficult to integrate without direct support. In future iterations, the proposed architecture, including extensions, could be implemented in a production web server. However, this responsibility is outside the scope of this research project and lies with the curators and creators of the data.

During the workshop with domain experts, another limitation was discussed: provenance. Data-altering improvements can also be captured in the artifact. If these changes are stored in the knowledge graph, other researchers could backtrack and understand the changes made in the data, thus improving the transparency of the artifact. However, provenance information should be stored in a separate data layer to avoid graph pollution.

The developed artifact, a knowledge graph, aims to support historians and other users in their research. While the methodology also aims to develop the artifact sustainably, such that the artifact can be used and expanded over the years, we can conclude that the artifact can be deemed usable, as illustrated by Table \ref{tab:competencyquestions}. Close to all competency questions could be answered by using the knowledge graph. Due to the unavailability of data regarding events that had an impact on a voyage but did not on the ship, the last competency question ('Can external factors, such as war, be linked to the intensity of shipping?') could not be answered. Such research questions provide a useful insight into the types of research questions that domain experts are interested in and a solid direction for future extensions of the model and knowledge graph. 

Furthermore, the participants of the workshop further confirmed the usability. A sidestep has been made by evaluating the artifact with the data story. During this evaluation, it was apparent that the artifact was useful for a broader range of people even though only a single evaluation was performed.

The results regarding the sustainability of the methods used and the developed artifact are multifaceted. There is no single answer that defines a sustainable artifact and process. In the related project section, we discussed the sustainability of the approaches taken by earlier project. We used these insights to shape our project in a more sustainable way.

Developing a sustainable knowledge graph consists of making design decisions. On a high or detailed level, design decisions influence the sustainability aspects. Picking between these design options can often lead to dilemmas where there is no clear answer. The choice between data usability and data correctness is one of these dilemmas. Do the data need to be accurate, even if this means the data will be difficult to process? In this project, we chose to store both original and cleaned dates. We faced another dilemma during the first review period. In this review, we examined the utilization of CIDOC CRM. The intention to re-use existing concepts of an ontology can conflict with the formal definition or use-case of that ontology. This in turn diminishes the sustainability as different projects might use the concept differently.

The two design cycles make the project stand out as the development of the artifact benefits immensely from receiving feedback. Design decisions made during the cycles are dependent on the researchers involved. Personal beliefs, experience, and other factors might steer the artifact in a different direction. Therefore, the transparent and accessible nature of the project is of utmost importance. It means that other researchers can not only continue with further iterations but also backtrack decisions and take a different path. Extended documentation, along with  code an a Docker instance, can be found on GitHub.\footnote{\url{https://github.com/stijnschout3n/thewindinoursails}}

\begin{acks}
The authors want to thank Francesco Beretta, Vincent Alamerc\-erey for help with CIDOC CRM and participants in the workshop of KNAW HuC, the Huygens Institute and the University of Amsterdam. 
\end{acks}

\bibliographystyle{ACM-Reference-Format}
\balance
\bibliography{thesis_ref}


\begin{thebibliography}{00}


\ifx \showCODEN    \undefined \def \showCODEN     #1{\unskip}     \fi
\ifx \showDOI      \undefined \def \showDOI       #1{#1}\fi
\ifx \showISBNx    \undefined \def \showISBNx     #1{\unskip}     \fi
\ifx \showISBNxiii \undefined \def \showISBNxiii  #1{\unskip}     \fi
\ifx \showISSN     \undefined \def \showISSN      #1{\unskip}     \fi
\ifx \showLCCN     \undefined \def \showLCCN      #1{\unskip}     \fi
\ifx \shownote     \undefined \def \shownote      #1{#1}          \fi
\ifx \showarticletitle \undefined \def \showarticletitle #1{#1}   \fi
\ifx \showURL      \undefined \def \showURL       {\relax}        \fi
\providecommand\bibfield[2]{#2}
\providecommand\bibinfo[2]{#2}
\providecommand\natexlab[1]{#1}
\providecommand\showeprint[2][]{arXiv:#2}

\bibitem[\protect\citeauthoryear{Azzaoui, Jacoby, Senger, Rodríguez, Loza,
  Zdrazil, Pinto, Williams, de~la Torre, Mestres, Pastor, Taboureau, Rarey,
  Chichester, Pettifer, Blomberg, Harland, Williams-Jones, and Ecker}{Azzaoui
  et~al\mbox{.}}{2013}]%
        {azzaoui_scientific_2013}
\bibfield{author}{\bibinfo{person}{Kamal Azzaoui}, \bibinfo{person}{Edgar
  Jacoby}, \bibinfo{person}{Stefan Senger}, \bibinfo{person}{Emiliano~Cuadrado
  Rodríguez}, \bibinfo{person}{Mabel Loza}, \bibinfo{person}{Barbara Zdrazil},
  \bibinfo{person}{Marta Pinto}, \bibinfo{person}{Antony~J. Williams},
  \bibinfo{person}{Victor de~la Torre}, \bibinfo{person}{Jordi Mestres},
  \bibinfo{person}{Manuel Pastor}, \bibinfo{person}{Olivier Taboureau},
  \bibinfo{person}{Matthias Rarey}, \bibinfo{person}{Christine Chichester},
  \bibinfo{person}{Steve Pettifer}, \bibinfo{person}{Niklas Blomberg},
  \bibinfo{person}{Lee Harland}, \bibinfo{person}{Bryn Williams-Jones}, {and}
  \bibinfo{person}{Gerhard~F. Ecker}.} \bibinfo{year}{2013}\natexlab{}.
\newblock \showarticletitle{Scientific competency questions as the basis for
  semantically enriched open pharmacological space development}.
\newblock \bibinfo{journal}{{\em Drug Discovery Today\/}} \bibinfo{volume}{18},
  \bibinfo{number}{17-18} (\bibinfo{date}{Sept.} \bibinfo{year}{2013}),
  \bibinfo{pages}{843--852}.
\newblock
\showISSN{13596446}
\showDOI{%
\url{https://doi.org/10.1016/j.drudis.2013.05.008}}


\bibitem[\protect\citeauthoryear{Beretta, Alamercery, Derks, Petram, and
  Schneider}{Beretta et~al\mbox{.}}{2019}]%
        {beretta1}
\bibfield{author}{\bibinfo{person}{Francesco Beretta}, \bibinfo{person}{Vincent
  Alamercery}, \bibinfo{person}{Sebastiaan Derks}, \bibinfo{person}{Lodewijk
  Petram}, {and} \bibinfo{person}{Jonas Schneider}.}
  \bibinfo{year}{2019}\natexlab{}.
\newblock \bibinfo{title}{{Geohistorical FAIR data: data integration and
  Interoperability using the OntoME platform}}.
\newblock \bibinfo{howpublished}{{Time Machine Conference 2019}}.
  (\bibinfo{date}{Oct.} \bibinfo{year}{2019}).
\newblock
\showURL{%
\url{https://halshs.archives-ouvertes.fr/halshs-02314003}}
\newblock
\shownote{Poster.}


\bibitem[\protect\citeauthoryear{Berry}{Berry}{2012}]%
        {berry_understanding_2012}
\bibfield{editor}{\bibinfo{person}{David~M. Berry}} (Ed.).
  \bibinfo{year}{2012}\natexlab{}.
\newblock \bibinfo{booktitle}{{\em Understanding digital humanities}}.
\newblock \bibinfo{publisher}{Palgrave Macmillan},
  \bibinfo{address}{Houndmills, Basingstoke, Hampshire ; New York}.
\newblock
\showISBNx{9780230292659}
\newblock
\shownote{OCLC: ocn701020028.}


\bibitem[\protect\citeauthoryear{Blanke, Bodard, Bryant, Dunn, Hedges, Jackson,
  and Scott}{Blanke et~al\mbox{.}}{2012}]%
        {blanke2012linked}
\bibfield{author}{\bibinfo{person}{Tobias Blanke}, \bibinfo{person}{Gabriel
  Bodard}, \bibinfo{person}{Michael Bryant}, \bibinfo{person}{Stuart Dunn},
  \bibinfo{person}{Mark Hedges}, \bibinfo{person}{Michael Jackson}, {and}
  \bibinfo{person}{David Scott}.} \bibinfo{year}{2012}\natexlab{}.
\newblock \showarticletitle{Linked data for humanities research—The SPQR
  experiment}. In \bibinfo{booktitle}{{\em 2012 6th IEEE international
  conference on digital ecosystems and technologies (DEST)}}.
  \bibinfo{publisher}{IEEE}, \bibinfo{pages}{1--6}.
\newblock


\bibitem[\protect\citeauthoryear{Christen}{Christen}{2012}]%
        {christen_data_2012}
\bibfield{author}{\bibinfo{person}{Peter Christen}.}
  \bibinfo{year}{2012}\natexlab{}.
\newblock \bibinfo{booktitle}{{\em Data matching: concepts and techniques for
  record linkage, entity resolution, and duplicate detection}}.
\newblock \bibinfo{publisher}{Springer}, \bibinfo{address}{Berlin ; New York}.
\newblock
\showISBNx{9783642311635}
\newblock
\shownote{OCLC: ocn809643173.}


\bibitem[\protect\citeauthoryear{Cidoc}{Cidoc}{2003}]%
        {cidoc2003cidoc}
\bibfield{author}{\bibinfo{person}{Crm Cidoc}.}
  \bibinfo{year}{2003}\natexlab{}.
\newblock \bibinfo{title}{The CIDOC conceptual reference model}.
\newblock   (\bibinfo{year}{2003}).
\newblock


\bibitem[\protect\citeauthoryear{{Claire Warwick}, Terras, and Nyhan}{{Claire
  Warwick} et~al\mbox{.}}{2012}]%
        {claire_warwick_digital_2012}
\bibfield{editor}{\bibinfo{person}{{Claire Warwick}},
  \bibinfo{person}{Melissa~M. Terras}, {and} \bibinfo{person}{Julianne Nyhan}}
  (Eds.). \bibinfo{year}{2012}\natexlab{}.
\newblock \bibinfo{booktitle}{{\em Digital humanities in practice:
  {Digitisation} and digital resources in the humanities}}.
\newblock \bibinfo{publisher}{Facet Publishing in association with UCL Centre
  for Digital Humanities}, \bibinfo{address}{London}.
\newblock
\showISBNx{9781856047661}
\newblock
\shownote{OCLC: ocn707962666.}


\bibitem[\protect\citeauthoryear{de~Boer, van Rossum, and Hoekstra}{de~Boer
  et~al\mbox{.}}{2015}]%
        {de_boer_dutch_2015}
\bibfield{author}{\bibinfo{person}{Victor de Boer}, \bibinfo{person}{Matthias
  van Rossum}, {and} \bibinfo{person}{Rik Hoekstra}.}
  \bibinfo{year}{2015}\natexlab{}.
\newblock \showarticletitle{The {Dutch} {Ships} and {Sailors} {Project}}.
\newblock \bibinfo{journal}{{\em DHCommons Journal\/}}  \bibinfo{volume}{1}
  (\bibinfo{date}{July} \bibinfo{year}{2015}).
\newblock


\bibitem[\protect\citeauthoryear{Dimou}{Dimou}{2020}]%
        {janev_knowledge_2020}
\bibfield{author}{\bibinfo{person}{A. Dimou}.} \bibinfo{year}{2020}\natexlab{}.
\newblock \bibinfo{booktitle}{{\em Knowledge {Graphs} and {Big} {Data}
  {Processing}}}. \bibinfo{series}{Lecture {Notes} in {Computer} {Science}},
  Vol.~\bibinfo{volume}{12072}.
\newblock \bibinfo{publisher}{Springer International Publishing},
  \bibinfo{address}{Cham}.
\newblock
\showISBNx{9783030531980}
\showDOI{%
\url{https://doi.org/10.1007/978-3-030-53199-7}}


\bibitem[\protect\citeauthoryear{Entjes}{Entjes}{2015}]%
        {entjes_linking_2015}
\bibfield{author}{\bibinfo{person}{Jeroen Entjes}.}
  \bibinfo{year}{2015}\natexlab{}.
\newblock {\em \bibinfo{title}{Linking {Maritime} {Datasets} to {Dutch} {Ships}
  and {Sailors} {Cloud} - {Case} studies on {Archangelvaart} and {Elbing}}}.
\newblock \bibinfo{thesistype}{Ph.D. Dissertation}. \bibinfo{school}{VU
  Amsterdam}.
\newblock
\showURL{%
\url{http://www.victordeboer.com/wp-content/uploads/2015/08/jeroen_entjes_final_thesis.pdf}}


\bibitem[\protect\citeauthoryear{Gelderblom, de~Jong, and Jonker}{Gelderblom
  et~al\mbox{.}}{2013}]%
        {gelderblom_formative_2013}
\bibfield{author}{\bibinfo{person}{Oscar Gelderblom}, \bibinfo{person}{Abe de
  Jong}, {and} \bibinfo{person}{Joost Jonker}.}
  \bibinfo{year}{2013}\natexlab{}.
\newblock \showarticletitle{The {Formative} {Years} of the {Modern}
  {Corporation}: {The} {Dutch} {East} {India} {Company} {VOC}, 1602–1623}.
\newblock \bibinfo{journal}{{\em The Journal of Economic History\/}}
  \bibinfo{volume}{73}, \bibinfo{number}{4} (\bibinfo{date}{Dec.}
  \bibinfo{year}{2013}), \bibinfo{pages}{1050--1076}.
\newblock
\showISSN{0022-0507, 1471-6372}
\showDOI{%
\url{https://doi.org/10.1017/S0022050713000879}}


\bibitem[\protect\citeauthoryear{Haslhofer, Isaac, and Simon}{Haslhofer
  et~al\mbox{.}}{2018}]%
        {haslhofer_knowledge_2018}
\bibfield{author}{\bibinfo{person}{Bernhard Haslhofer},
  \bibinfo{person}{Antoine Isaac}, {and} \bibinfo{person}{Rainer Simon}.}
  \bibinfo{year}{2018}\natexlab{}.
\newblock \showarticletitle{Knowledge {Graphs} in the {Libraries} and {Digital}
  {Humanities} {Domain}}.
\newblock \bibinfo{journal}{{\em arXiv:1803.03198 [cs]\/}}
  (\bibinfo{year}{2018}), \bibinfo{pages}{1--8}.
\newblock
\showDOI{%
\url{https://doi.org/10.1007/978-3-319-63962-8_291-1}}
\newblock
\shownote{arXiv: 1803.03198.}


\bibitem[\protect\citeauthoryear{Hendriks, Groth, and van Erp}{Hendriks
  et~al\mbox{.}}{2020}]%
        {hendriks_recognising_2020}
\bibfield{author}{\bibinfo{person}{B Hendriks}, \bibinfo{person}{P Groth},
  {and} \bibinfo{person}{M van Erp}.} \bibinfo{year}{2020}\natexlab{}.
\newblock \showarticletitle{Recognising and linking entities in old dutchtext:
  a case study on voc notary records}.
\newblock \bibinfo{journal}{{\em Collect \& Connect Leiden\/}}
  (\bibinfo{year}{2020}).
\newblock


\bibitem[\protect\citeauthoryear{{Hevner}, {March}, {Park}, and {Ram}}{{Hevner}
  et~al\mbox{.}}{2004}]%
        {hevner_design_2004}
\bibfield{author}{\bibinfo{person}{{Hevner}}, \bibinfo{person}{{March}},
  \bibinfo{person}{{Park}}, {and} \bibinfo{person}{{Ram}}.}
  \bibinfo{year}{2004}\natexlab{}.
\newblock \showarticletitle{Design {Science} in {Information} {Systems}
  {Research}}.
\newblock \bibinfo{journal}{{\em MIS Quarterly\/}} \bibinfo{volume}{28},
  \bibinfo{number}{1} (\bibinfo{year}{2004}), \bibinfo{pages}{75}.
\newblock
\showISSN{02767783}
\showDOI{%
\url{https://doi.org/10.2307/25148625}}


\bibitem[\protect\citeauthoryear{Hughes}{Hughes}{2012}]%
        {hughes_evaluating_2012}
\bibfield{editor}{\bibinfo{person}{Lorna~M. Hughes}} (Ed.).
  \bibinfo{year}{2012}\natexlab{}.
\newblock \bibinfo{booktitle}{{\em Evaluating and measuring the value, use and
  impact of digital collections}}.
\newblock \bibinfo{publisher}{Facet Publ}, \bibinfo{address}{London}.
\newblock
\showISBNx{9781856047203}
\newblock
\shownote{OCLC: 774166752.}


\bibitem[\protect\citeauthoryear{Hyv{\"o}nen et~al\mbox{.}}{Hyv{\"o}nen
  et~al\mbox{.}}{2020}]%
        {hyvonen2020linked}
\bibfield{author}{\bibinfo{person}{Eero Hyv{\"o}nen} {et~al\mbox{.}}}
  \bibinfo{year}{2020}\natexlab{}.
\newblock \showarticletitle{Linked open data infrastructure for digital
  humanities in finland}.
\newblock \bibinfo{journal}{{\em Proceedings of Digital Humanities in Nordic
  Countries (DHN 2020)\/}} (\bibinfo{year}{2020}).
\newblock


\bibitem[\protect\citeauthoryear{Jovanovik}{Jovanovik}{2016}]%
        {Jovanovik}
\bibfield{author}{\bibinfo{person}{Milos Jovanovik}.}
  \bibinfo{year}{2016}\natexlab{}.
\newblock {\em \bibinfo{title}{Linked Data Application Development
  Methodology}}.
\newblock \bibinfo{thesistype}{Ph.D. Dissertation}.
\newblock


\bibitem[\protect\citeauthoryear{Petram, Koolen, van Koert, Wevers, and van
  Lottum}{Petram et~al\mbox{.}}{}]%
        {petram_data_nodate}
\bibfield{author}{\bibinfo{person}{Lodewijk Petram}, \bibinfo{person}{Marijn
  Koolen}, \bibinfo{person}{Rutger van Koert}, \bibinfo{person}{Melvin Wevers},
  {and} \bibinfo{person}{Jelle van Lottum}.}
\newblock \showarticletitle{Data on the {Maritime} {Workforce} of the {Dutch}
  {East} {India} {Company} in the 18th {Century}}.
\newblock \bibinfo{journal}{{\em to be published\/}} (\bibinfo{year}{????}).
\newblock


\bibitem[\protect\citeauthoryear{Waagmeester, Stupp, Burgstaller-Muehlbacher,
  Good, Griffith, Griffith, Hanspers, Hermjakob, Hudson, Hybiske, Keating,
  Manske, Mayers, Mietchen, Mitraka, Pico, Putman, Riutta, Queralt-Rosinach,
  Schriml, Shafee, Slenter, Stephan, Thornton, Tsueng, Tu, Ul-Hasan,
  Willighagen, Wu, and Su}{Waagmeester et~al\mbox{.}}{2020}]%
        {waagmeester_wikidata_2020}
\bibfield{author}{\bibinfo{person}{Andra Waagmeester}, \bibinfo{person}{Gregory
  Stupp}, \bibinfo{person}{Sebastian Burgstaller-Muehlbacher},
  \bibinfo{person}{Benjamin~M Good}, \bibinfo{person}{Malachi Griffith},
  \bibinfo{person}{Obi~L Griffith}, \bibinfo{person}{Kristina Hanspers},
  \bibinfo{person}{Henning Hermjakob}, \bibinfo{person}{Toby~S Hudson},
  \bibinfo{person}{Kevin Hybiske}, \bibinfo{person}{Sarah~M Keating},
  \bibinfo{person}{Magnus Manske}, \bibinfo{person}{Michael Mayers},
  \bibinfo{person}{Daniel Mietchen}, \bibinfo{person}{Elvira Mitraka},
  \bibinfo{person}{Alexander~R Pico}, \bibinfo{person}{Timothy Putman},
  \bibinfo{person}{Anders Riutta}, \bibinfo{person}{Nuria Queralt-Rosinach},
  \bibinfo{person}{Lynn~M Schriml}, \bibinfo{person}{Thomas Shafee},
  \bibinfo{person}{Denise Slenter}, \bibinfo{person}{Ralf Stephan},
  \bibinfo{person}{Katherine Thornton}, \bibinfo{person}{Ginger Tsueng},
  \bibinfo{person}{Roger Tu}, \bibinfo{person}{Sabah Ul-Hasan},
  \bibinfo{person}{Egon Willighagen}, \bibinfo{person}{Chunlei Wu}, {and}
  \bibinfo{person}{Andrew~I Su}.} \bibinfo{year}{2020}\natexlab{}.
\newblock \showarticletitle{Wikidata as a knowledge graph for the life
  sciences}.
\newblock \bibinfo{journal}{{\em eLife\/}}  \bibinfo{volume}{9}
  (\bibinfo{date}{March} \bibinfo{year}{2020}), \bibinfo{pages}{e52614}.
\newblock
\showISSN{2050-084X}
\showDOI{%
\url{https://doi.org/10.7554/eLife.52614}}


\bibitem[\protect\citeauthoryear{Wieringa}{Wieringa}{2014}]%
        {wieringa2014design}
\bibfield{author}{\bibinfo{person}{Roel~J Wieringa}.}
  \bibinfo{year}{2014}\natexlab{}.
\newblock \bibinfo{booktitle}{{\em Design science methodology for information
  systems and software engineering}}.
\newblock \bibinfo{publisher}{Springer}.
\newblock


\bibitem[\protect\citeauthoryear{Wilkinson, Dumontier, Aalbersberg, Appleton,
  Axton, Baak, Blomberg, Boiten, da~Silva~Santos, Bourne, Bouwman, Brookes,
  Clark, Crosas, Dillo, Dumon, Edmunds, Evelo, Finkers, Gonzalez-Beltran, Gray,
  Groth, Goble, Grethe, Heringa, ’t Hoen, Hooft, Kuhn, Kok, Kok, Lusher,
  Martone, Mons, Packer, Persson, Rocca-Serra, Roos, van Schaik, Sansone,
  Schultes, Sengstag, Slater, Strawn, Swertz, Thompson, van~der Lei, van
  Mulligen, Velterop, Waagmeester, Wittenburg, Wolstencroft, Zhao, and
  Mons}{Wilkinson et~al\mbox{.}}{2016}]%
        {wilkinson_fair_2016}
\bibfield{author}{\bibinfo{person}{Mark~D. Wilkinson}, \bibinfo{person}{Michel
  Dumontier}, \bibinfo{person}{IJsbrand~Jan Aalbersberg},
  \bibinfo{person}{Gabrielle Appleton}, \bibinfo{person}{Myles Axton},
  \bibinfo{person}{Arie Baak}, \bibinfo{person}{Niklas Blomberg},
  \bibinfo{person}{Jan-Willem Boiten}, \bibinfo{person}{Luiz~Bonino da
  Silva~Santos}, \bibinfo{person}{Philip~E. Bourne}, \bibinfo{person}{Jildau
  Bouwman}, \bibinfo{person}{Anthony~J. Brookes}, \bibinfo{person}{Tim Clark},
  \bibinfo{person}{Mercè Crosas}, \bibinfo{person}{Ingrid Dillo},
  \bibinfo{person}{Olivier Dumon}, \bibinfo{person}{Scott Edmunds},
  \bibinfo{person}{Chris~T. Evelo}, \bibinfo{person}{Richard Finkers},
  \bibinfo{person}{Alejandra Gonzalez-Beltran}, \bibinfo{person}{Alasdair~J.G.
  Gray}, \bibinfo{person}{Paul Groth}, \bibinfo{person}{Carole Goble},
  \bibinfo{person}{Jeffrey~S. Grethe}, \bibinfo{person}{Jaap Heringa},
  \bibinfo{person}{Peter~A.C ’t Hoen}, \bibinfo{person}{Rob Hooft},
  \bibinfo{person}{Tobias Kuhn}, \bibinfo{person}{Ruben Kok},
  \bibinfo{person}{Joost Kok}, \bibinfo{person}{Scott~J. Lusher},
  \bibinfo{person}{Maryann~E. Martone}, \bibinfo{person}{Albert Mons},
  \bibinfo{person}{Abel~L. Packer}, \bibinfo{person}{Bengt Persson},
  \bibinfo{person}{Philippe Rocca-Serra}, \bibinfo{person}{Marco Roos},
  \bibinfo{person}{Rene van Schaik}, \bibinfo{person}{Susanna-Assunta Sansone},
  \bibinfo{person}{Erik Schultes}, \bibinfo{person}{Thierry Sengstag},
  \bibinfo{person}{Ted Slater}, \bibinfo{person}{George Strawn},
  \bibinfo{person}{Morris~A. Swertz}, \bibinfo{person}{Mark Thompson},
  \bibinfo{person}{Johan van~der Lei}, \bibinfo{person}{Erik van Mulligen},
  \bibinfo{person}{Jan Velterop}, \bibinfo{person}{Andra Waagmeester},
  \bibinfo{person}{Peter Wittenburg}, \bibinfo{person}{Katherine Wolstencroft},
  \bibinfo{person}{Jun Zhao}, {and} \bibinfo{person}{Barend Mons}.}
  \bibinfo{year}{2016}\natexlab{}.
\newblock \showarticletitle{The {FAIR} {Guiding} {Principles} for scientific
  data management and stewardship}.
\newblock \bibinfo{journal}{{\em Scientific Data\/}} \bibinfo{volume}{3},
  \bibinfo{number}{1} (\bibinfo{date}{Dec.} \bibinfo{year}{2016}),
  \bibinfo{pages}{160018}.
\newblock
\showISSN{2052-4463}
\showDOI{%
\url{https://doi.org/10.1038/sdata.2016.18}}


\end{thebibliography}
 
\end{document}